\documentclass[12pt,preprint]{aastex}
\begin{document}
\title{A Cosmological Mass Function with Broken Hierarchy}
\author{\sc Jounghun Lee}
\affil{School of Physics and Astronomy, FPRD, Seoul National University, 
Seoul 151-742, Korea}
\email{jounghun@astro.snu.ac.kr}

\begin{abstract}
We construct an analytic formalism for the mass function of cold dark matter 
halos, assuming that there is a break in the hierarchical merging process.  
According to this {\it broken-hierarchy} scenario, due to the inherent
nature of the gravitational tidal field the formation of massive pancakes 
precedes that of dark halos of low-mass. In the framework of the Zel'dovich 
approximation which generically predicts the presence of pancakes, we first 
derive analytically the conditional probability that a low-mass halo observed 
at present epoch was embedded in an isolated pancake at some earlier epoch. 
Then, we follow the standard Press-Schechter approach to count analytically 
the number density of low-mass halos that formed through anti-hierarchical 
fragmentation of the massive pancakes. Our mass function is well approximated 
by a power-law $dN/dM = M^{-l}$ in the mass range 
$10^{6}h^{-1}M_{\odot}\le M\le 10^{10}h^{-1}M_{\odot}$ with the slope $l=1.86$ 
shallower than that of the currently popular Sheth-Tormen mass function 
$l = 2.1$. It is expected that our mass function will provide a useful 
analytic tool for investigating the effect of broken hierarchy on the 
structure formation.
\end{abstract} 

\section{INTRODUCTION}

In the cold dark matter (CDM) paradigm the gravitationally bound objects 
made of dark matter particles (dark halos) are supposed to form through 
hierarchical merging process.  \citet[][hereafter PS]{pre-sch74} 
devised for the first time an analytic formalism to evaluate the mass 
distribution function of CDM halos that formed hierarchically. Later 
\citet[][hereafter ST]{she-tor99} refined the PS mass function by 
taking into account the non-spherical collapse. The ST mass function 
has been tested against many N-body simulations, showing good agreements 
\citep[e.g.,][]{ree-etal03}. This success of the ST mass function implies 
the validity of the hierarchical merging scenario since it was originally 
derived under the assumption that the CDM halos form through hierarchical 
merging. 

Yet it is premature to assert that the formation of CDM halos is always 
hierarchical over the entire mass range given the fact that the validity of 
the ST mass function was rather limited to the relatively high- mass section 
($>10^{10}h^{-1}M_{\odot}$). Results from N-body simulations still suffer 
from the lack of information on the low-mass section 
($\le 10^{10}h^{-1}M_{\odot}$) due to the resolution limit. 

Very recently, \citet{mo-etal05} came up with a new halo-formation scenario 
where the formation of low-mass halos is somewhat anti-hierarchical, preceded 
by that of massive pancakes. According to their model, the preheated medium 
caused by the gravitational pancaking effect suppresses the star-formation 
sufficiently in the low-mass halos, which can explain the observed low HI-mass 
as well as the faint-end slope of the galaxy luminosity function.  

Although the work of \citet{mo-etal05} was focused on explaining the 
suppression of star-formation in the low-mass halos, we note that if the 
formation of low-mass halos was indeed preceded by the formation of massive 
pancakes, what is suppressed should not be only the star-formation but also 
the formation of low-mass halos itself. To take into account the {\it break} 
in the hierarchical process and to predict the abundance of low-mass halos 
more accurately, it is desirable to have an analytic model for it derived 
from first principles.

Our goal here is to construct an analytic model for the low-mass halos that 
form through anti-hierarchical fragmentation of massive pancakes. 
To achieve this goal, we adopt the Zel'dovich approximation as a simplest 
footstep which generically predicts the formation of pancakes, and we follow  
the standard PS approach to count the abundance of dark halos as a function 
of mass. The hypotheses, the mathematical layout and the predictions 
of our model are presented in $\S 2$, and the summary and discussion of the 
final results are provided in $\S 3$. 

\section{ANALYTIC FORMALISM}

In the Press-Schecther formalism \citep[][hereafter, PS]{pre-sch74}, an 
{\it isolated} dark halo (a halo just collapsed) of mass $M$ forms from 
the regions in the density field whose average density contrast $\delta \equiv 
\Delta\rho/\bar{\rho}$ ($\bar{\rho}$: the mean mass density of the universe) 
smoothed on the mass scale of $M$ reaches some critical value, $\delta_{c}$. 
The value of the critical density $\delta_{c}$ is approximately $1.68$, 
which depends very weakly on the background cosmology and the redshift 
\citep{kit-sut96}.

The Gaussian probability distribution of the linear density contrast smoothed 
with the sharp k-space filter on the mass scale $M$ is given as  
\begin{equation}
\label{eqn:den}
p(\delta) = \frac{1}{\sqrt{2\pi}\sigma}
\exp\left[-\frac{\delta^{2}}{2\sigma^{2}}\right], \quad 
\sigma^{2}(M) \equiv \int^{\ln k_{c}}_{-\infty}\!\!\Delta^{2}(k)\,d\ln k,
\quad M = 6\pi^{2}\bar{\rho}\,k^{-3}_{c}, 
\end{equation}
where $\sigma(M)$ is the rms density fluctuation, and $\Delta^{2}(k)$ is the 
dimensionless power spectrum. Throughout this Letter, we use the power 
spectrum of the concordance $\Lambda$CDM cosmology with $\Omega_{\Lambda}=0.7,
\Omega_{m}=0.3,\Omega_{b}=0.044,h=0.7$ \citep{bar-etal86}.

In the Zel'dovich approximation \citep[][hereafter, ZEL]{zel70}, the mass 
density is given as
\begin{equation}
\label{eqn:zel}
\rho = \frac{\bar{\rho}}{(1-\lambda_{1})(1-\lambda_{2})(1-\lambda_{1})}, 
\end{equation}
where $\lambda_{1},\lambda_{2},\lambda_{3}$ (with $\lambda_{1} \ge 
\lambda_{2} \ge \lambda_{3}$) are the three eigenvalues of the deformation 
tensor $d_{ij}$ defined as the second derivative of the perturbation potential 
$\Phi$: $d_{ij} \equiv \partial_{i}\partial_{j}\Phi$. 
\citet{dor70} derived the joint distribution of the three eigenvalues:
\begin{equation}
\label{eqn:lam}
p(\lambda_{1},\lambda_{2},
\lambda_{3}) = \frac{3375}{8\sqrt{5}\pi\sigma^{6}}
\exp\left(-\frac{3I^{2}_{1}}{\sigma^{2}} + \frac{15I_{2}}{2\sigma^{2}}\right)
(\lambda_{1}-\lambda_{2})(\lambda_{2}-\lambda_{3})(\lambda_{1}-\lambda_{3}),
\end{equation}
with $I_{1} \equiv \lambda_{1} + \lambda_{2} + \lambda_{3}$ 
and $I_{2} \equiv \lambda_{1}\lambda_{2} + \lambda_{2}\lambda_{3} + 
\lambda_{2}\lambda_{3}$.

Equation (\ref{eqn:zel}) implies that the mass-density diverges along the 
direction of the major principal axis of the deformation tensor if the largest 
eigenvalue reaches unity and the intermediate and the smallest eigenvalues 
are less than zero. In other words, an isolated pancake (a two-dimensional 
object just collapsed only along the first principal axis) of mass $M$ forms 
if the following condition is satisfied: 
$\lambda_{1}=\lambda_{c},\lambda_{2}<0,\lambda_{3}<0$ with $\lambda_{c}=1$ 
on the mass scale $M$.

Now that the conditions for the formation of isolated halos and pancakes are 
specified, we would like to find the probability that a halo at present 
epoch was embedded in a pancake at some earlier epoch before it formed. 
For this, it is required to have the joint distribution of the linear density 
and the three eigenvalues of the deformation tensor on two different scales 
on two different epochs. Let $\delta$ be defined at present epoch on the 
galactic mass scale $M_{g}$, and let $\lambda_{1},\lambda_{2},\lambda_{3}$ 
be defined at some earlier epoch of redshift $z>0$ on the larger mass scale 
$M_{p} > M_{g}$. The rms density fluctuation at redshift $z$ on mass 
scale $M_{p}$ is given as $\sigma(M_{p},z)=b(z)\sigma(M_{p})$ where 
$b(z)$ is the growth factor of the linear density that is normalized to 
satisfy $b(0)=1$. Since the growth factor as well as the rms density 
fluctuation is a decreasing function of $z$, we have 
$\sigma(M_{p},z)<\sigma(M_{g})$. 
From here on, we use the notations of $\sigma^{\prime}$,$d^{\prime}_{ij}$,
$\lambda^{\prime}_{1},\lambda^{\prime}_{2},\lambda^{\prime}_{3}$ to represent 
the rms density fluctuation, the deformation tensor and its three eigenvalues 
at redshift $z$ on the mass scale $M_{p}$.   

To derive the joint distribution of $\delta$ and $\lambda^{\prime}_{1},
\lambda^{\prime}_{2},\lambda^{\prime}_{3}$,  we first derive the multivariate 
Gaussian distribution of $\delta$ and the 6 independent components of the 
symmetric tensor, $d^{\prime}_{ij}$.
Rotating the frame into the principal axis of $d^{\prime}_{ij}$ and using 
the fact that $\delta$ is invariant under the axis-rotation, we derive 
analytically the joint distribution of $\delta$ and 
$\lambda^{\prime}_{1},\lambda^{\prime}_{2},\lambda^{\prime}_{3}$
\begin{eqnarray}
\label{eqn:ess}
p(\delta,\lambda^{\prime}_{1},\lambda^{\prime}_{2},
\lambda^{\prime}_{3}) &=& 
\frac{1}{\sqrt{2\pi}\sigma_{\Delta}}\frac{3375}{8\sqrt{5}\pi\sigma^{\prime6}}
\exp\left[-\frac{(\delta-I^{\prime}_{1})^{2}}
{2\sigma^{2}_{\Delta}}\right]\times \nonumber \\
&&\exp\left(-\frac{3I^{\prime 2}_{1}}{\sigma^{\prime 2}} + 
\frac{15I^{\prime}_{2}}{2\sigma^{\prime 2}}\right)
(\lambda^{\prime}_{1}-\lambda^{\prime}_{2})
(\lambda^{\prime}_{2}-\lambda^{\prime}_{3})
(\lambda^{\prime}_{1}-\lambda^{\prime}_{3}),
\end{eqnarray}
where $\sigma^{2}_{\Delta} \equiv \sigma^{2} - \sigma^{\prime 2}$, 
$I^{\prime}_{1}=\lambda^{\prime}_{1} + \lambda^{\prime}_{2} + 
\lambda^{\prime}_{3}$, and $I^{\prime}_{1}=\lambda^{\prime}_{1}
\lambda^{\prime}_{2}+\lambda^{\prime}_{2}\lambda^{\prime}_{3}+ 
\lambda^{\prime}_{1}\lambda^{\prime}_{3}$.

Through equations and (\ref{eqn:den}) and (\ref{eqn:ess}) we find the 
probability that a halo of mass $M_{g}$ observed at present epoch was 
embedded in an isolated pancake of larger mass $M_{p}$ at redshift $z$ 
with the help of the Bayes theorem: 
\begin{eqnarray}
\label{eqn:con1}
p(\lambda^{\prime}_{1}=\lambda_{c},\lambda^{\prime}_{2}<0,
\lambda^{\prime}_{3}<0\vert\delta \ge \delta_{c}) &=& 
\frac{p(\delta \ge \delta_{c},\lambda^{\prime}_{1}=\lambda_{c},
\lambda^{\prime}_{2}<0,\lambda^{\prime}_{3}<0)}{P(\delta \ge \delta_{c})}, 
\nonumber \\
&=& \frac{\int^{0}_{\lambda^{\prime}_{3}}\!d\lambda^{\prime}_{2}
\!\int^{0}_{-\infty}\!d\lambda^{\prime}_{3}\!\int^{0}_{-\infty}
\!d\delta\,p(\delta,\lambda^{\prime}_{1}=\lambda_{c},\lambda^{\prime}_{2},
\lambda^{\prime}_{3})} {\int_{\delta_{c}}^{\infty}d\delta p(\delta)}.
\end{eqnarray}
In equation (\ref{eqn:con1}), the integration over $\delta$ can be readily 
evaluated 
\begin{eqnarray}
\int^{0}_{\lambda^{\prime}_{3}}\!d\lambda^{\prime}_{2}
\!\int^{0}_{-\infty}\!d\lambda^{\prime}_{3}\!\int^{0}_{-\infty}
\!d\delta\,p(\delta,\lambda_{c},\lambda^{\prime}_{2},
\lambda^{\prime}_{3}) &=& \frac{1}{2}
\int^{0}_{\lambda^{\prime}_{3}}\!d\lambda^{\prime}_{2}\!
\int^{0}_{-\infty}\!d\lambda^{\prime}_{3}\,{\rm erfc}
\left(\frac{\delta_{c}-I^{\prime}_{1}}{\sqrt{2}
\sigma_{\Delta}}\right)\times \nonumber \\ 
&& p(\lambda^{\prime}_{1}=\lambda_{c},
\lambda^{\prime}_{2},\lambda^{\prime}_{3}), \\
\int_{\delta_{c}}^{\infty}d\delta\,p(\delta) &=& \frac{1}{2}
{\rm erfc}\left(\frac{\delta_{c}}{\sqrt{2}\sigma}\right).
\end{eqnarray}
This probability (eq.[\ref{eqn:con1}]) will allow us to determine the 
most-likely epoch when the formation of pancakes precedes that of low-mass 
halos, and the typical mass scale for the formation of pancakes as well.

For comparison, we consider the probability that a halo of mass $M_{g}$ 
observed at present epoch just formed hierarchically at redshift $z$ 
which is approximately given as \citep{bow91,lac-col94}: 
\begin{eqnarray}
\label{eqn:con2}
p(\delta^{\prime\prime}=\delta_{c}\vert\delta\ge\delta_{c}) &=& 
\frac{ p(\delta\ge\delta_{c},\delta^{\prime\prime}=\delta_{c})}
{P(\delta \ge \delta_{c})}, \nonumber \\
&=& \frac{1}{\sqrt{2\pi\sigma^{\prime\prime}}}
\left[{\rm erfc}\left(\frac{\delta_{c}}{\sqrt{2}\sigma}\right)\right]^{-1}
\exp\left(-\frac{\delta^{2}_{c}}{2\sigma^{\prime\prime 2}}\right), 
\end{eqnarray}
where $\delta^{\prime\prime}$ represents the linear density on the mass 
scale $M_{g}$ at redshift $z$. The relative difference between 
$p(\lambda^{\prime}_{1}=\lambda_{c},\lambda^{\prime}_{2}<0,
\lambda^{\prime}_{3}<0\vert\delta\ge\delta_{c})$ and 
$p(\delta^{\prime}=\delta_{c}\vert\delta\ge\delta_{c})$ indicates how 
probable the anti-hierarchical formation of low-mass halos is at given epoch.  

\begin{figure}
\begin{center}
\plotone{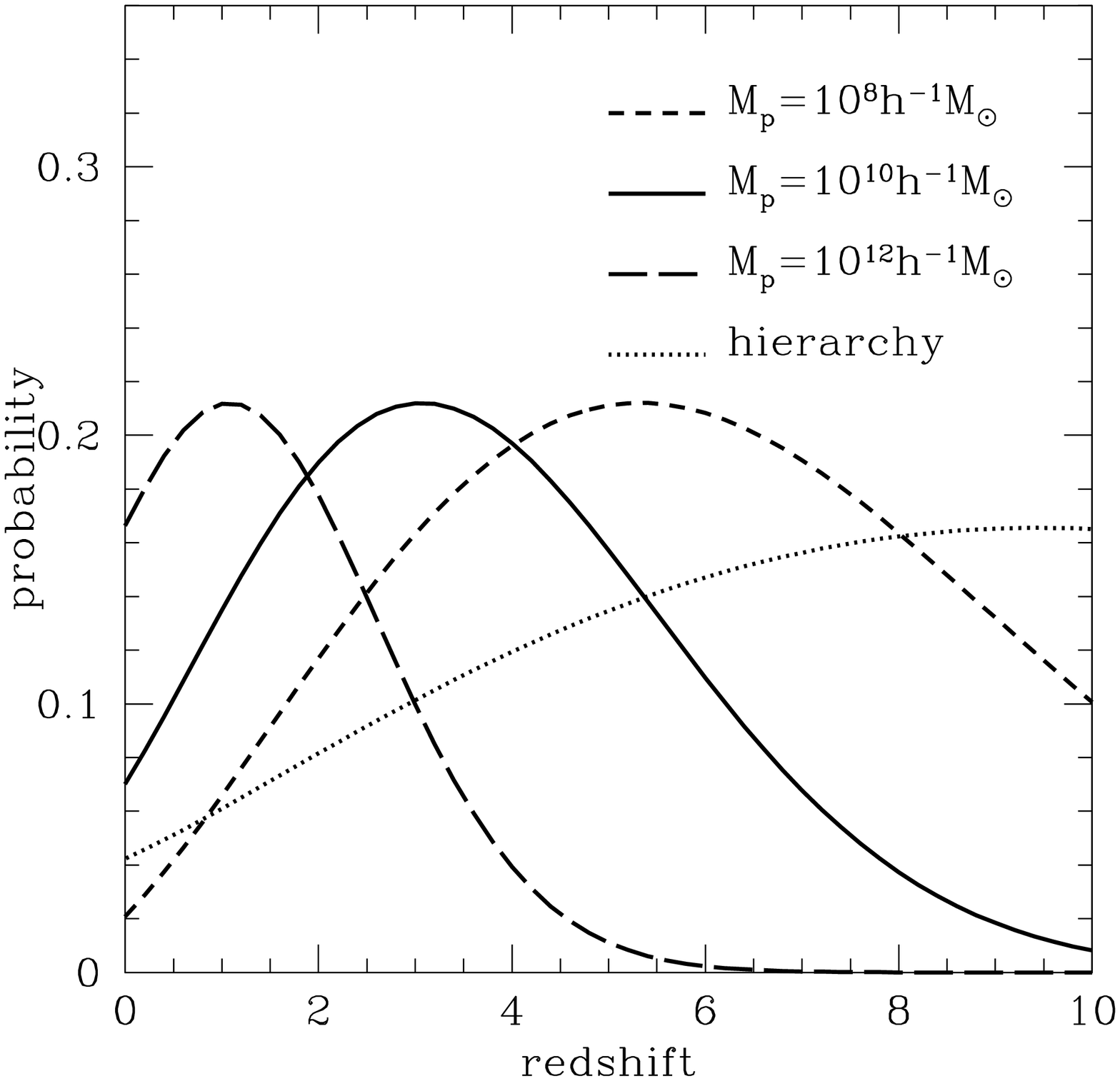}
\caption{Probability that a galactic halo observed on  mass scale, 
$M_{g} = 10^{6}h^{-1}M_{\odot}$ at present epoch was embedded in a 
pancake at redshift $z$ for the three cases of the pancake's mass: 
$M_{p} = 10^{8}h^{-1}M_{\odot}$ (dashed); $M_{p} = 10^{10}h^{-1}M_{\odot}$ 
(solid); $M_{p} = 10^{12}h^{-1}M_{\odot}$ (long dashed). For comparison, 
the probability that a galactic halo formed hierarchically at redshift $z$ 
is also plotted (dotted).
\label{fig:pan}}
\end{center}
\end{figure}

The direct comparison of the two conditional probabilities 
(eqs.[\ref{eqn:con1}] and [\ref{eqn:con2}]) is shown in Fig.\ref{fig:pan}. 
The halo mass $M_{g}$ observed at present epoch is set at the dwarf galactic 
scale $M_{g}=10^{6}h^{-1}M_{\odot}$, and three different cases of the 
pancake's mass $M_{p}$ are considered: 
$M_{p} = 10^{8}h^{-1}M_{\odot}$ (dashed); 
$M_{p}=10^{10}h^{-1}M_{\odot}$ (solid); 
$M_{p}=10^{12}h^{-1}M_{\odot}$ (long dashed). As can be seen, the value of 
$p(\lambda^{\prime}_{1}=\lambda_{c},\lambda^{\prime}_{2}<0,
\lambda^{\prime}_{3}<0\vert\delta\ge\delta_{c})$ is twice higher than that of 
$p(\delta^{\prime}=\delta_{c}\vert\delta\ge\delta_{c})$ at $z \sim 2$ 
for the case of $10^{10}h^{-1}M_{\odot}\le M_{p}\le 10^{12}h^{-1}M_{\odot}$.

\begin{figure}
\begin{center}
\plotone{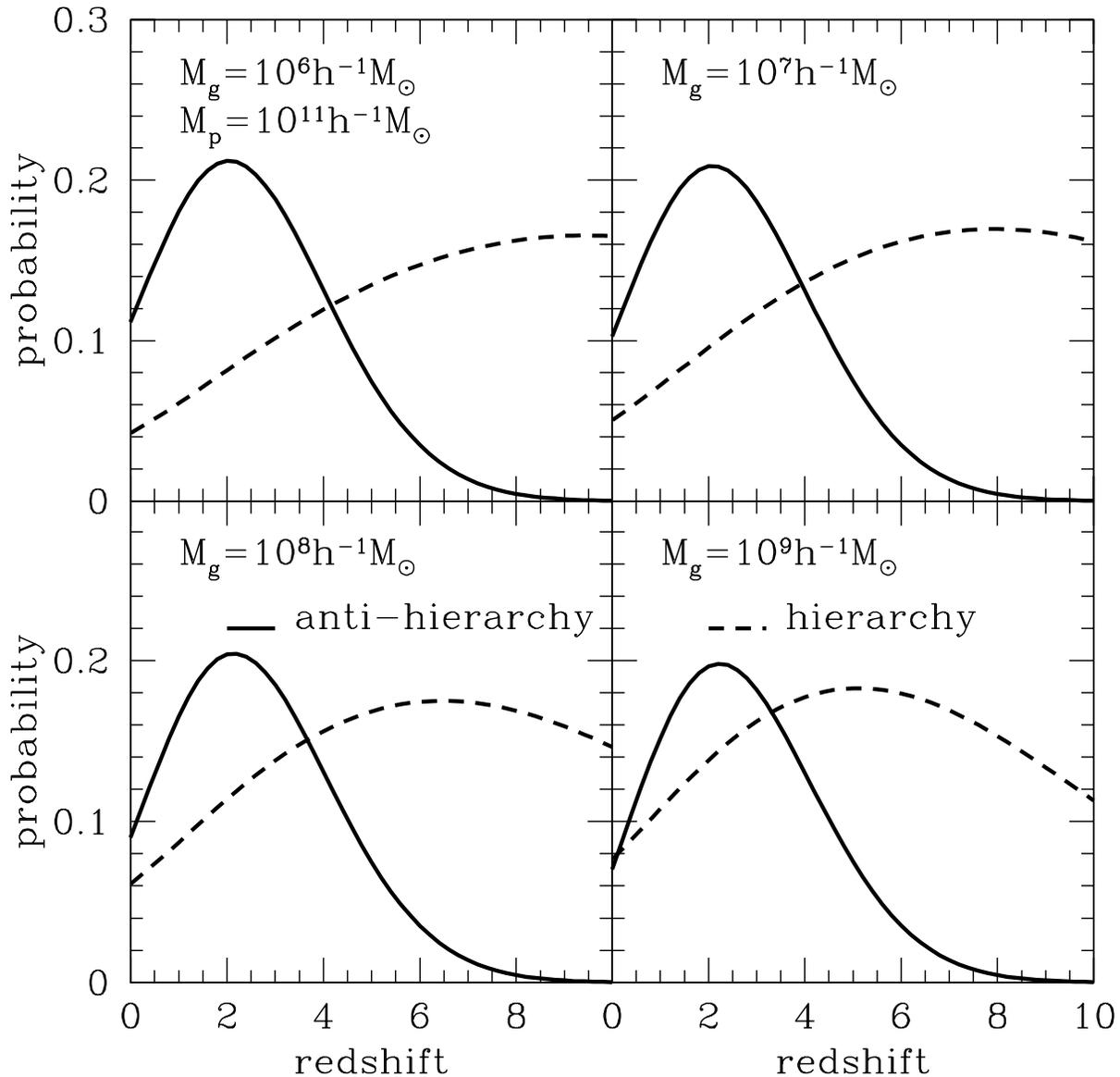}
\caption{Comparison of the probability that a galactic halo forms in the 
anti-hierarchically (solid) with the probability that it forms in the 
purely hierarchical way for four different cases of the halo mass 
$M_{g}$: $M_{g} = 10^{6}h^{-1}M_{\odot}$ (upper left); 
$M_{g} = 10^{7}h^{-1}M_{\odot}$ (upper right); 
$M_{g} = 10^{8}h^{-1}M_{\odot}$ (lower right);
$M_{g} = 10^{9}h^{-1}M_{\odot}$ (lower left). The pancake's mass is 
set at $M_{p}=10^{11}h^{-1}M_{\odot}$.
\label{fig:panel}}
\end{center}
\end{figure}

Figure \ref{fig:panel} also plots the two probabilities as solid and dashed 
lines. In this Fig. \ref{fig:panel} the pancake's mass is set at 
$M_{p}=10^{11}h^{-1}M_{\odot}$, and the four different cases of the halo mass 
$M_{g}$ are considered in separate panels:   
$M_{g}=10^{6}h^{-1}M_{\odot}$ (upper left); 
$M_{g}=10^{7}h^{-1}M_{\odot}$ (upper right);
$M_{g}=10^{8}h^{-1}M_{\odot}$ (lower left);
$M_{g}=10^{9}h^{-1}M_{\odot}$ (lower right).
As shown, the probability distribution 
$p(\lambda^{\prime}_{1}=\lambda_{c},\lambda^{\prime}_{2}<0,
\lambda^{\prime}_{3}<0\vert\delta\ge\delta_{c})$ has a maximum value around 
$z=2$, position of which shifts to the low-redshift section as the halo mass 
$M_{g}$ increases. For all four cases of $M_{g}$ at $z\sim 2$, 
the value of $p(\lambda^{\prime}_{1}=\lambda_{c},\lambda^{\prime}_{2}<0,
\lambda^{\prime}_{3}<0\vert\delta\ge\delta_{c})$ is consistently higher 
than that of $p(\delta^{\prime}=\delta_{c}\vert\delta\ge\delta_{c})$. 
The results shown in Figs. \ref{fig:pan} and \ref{fig:panel} imply that 
the halo of mass $M_{g}\le 10^{10}h^{-1}M_{\odot}$ observed at 
present epoch are more likely to have been embedded in massive 
pancakes of mass $M_{p} \approx 10^{11}h^{-1}M_{\odot}$ around $z=2$ 
rather than formed through hierarchical merging.

Setting the typical mass scale and redshift for the formation of pancakes at 
$10^{11}h^{-1}M_{\odot}$ and $z=2$, respectively, we follow the standard PS 
approach to evaluate the mass distribution function of the low-mass halos 
that formed anti-hierarchically. 
According to the theory the differential number density of the dark halos 
in the mass range $[M,M + dM]$ is related to the volume fraction 
$F$ occupied by the proto-halo regions in the linear density field that 
satisfy a specified collapse condition: 
\begin{equation}
\label{eqn:mf}
\frac{dN}{dM} \equiv A\frac{\bar{\rho}}{M}\left|\frac{dF}{dM}\right|
= A\frac{\bar{\rho}}{M^{2}}\left|\frac{d\sigma}{d\ln M}\right|
\left|\frac{\partial F}{\partial\sigma}\right|,
\end{equation}
where $A$ is the normalization factor, which is exactly $2$ in the original 
PS theory \citep{pea-hea90,bon-etal91,jed95}. 
If the halos observed at present epoch were embedded in massive 
pancakes at redshift $z$, the volume fraction $F$ should be written as
\begin{equation}
\label{eqn:vf}
F(\sigma)=\int_{\delta_{c}}^{\infty}\!\!d\delta\,
p(\delta\vert\lambda^{\prime}_{1}=\lambda_{c},\lambda^{\prime}_{2}< 0, 
\lambda^{\prime}_{3}<0).
\end{equation}
The conditional probability $p(\delta\vert\lambda^{\prime}_{1}=\lambda_{c},
\lambda^{\prime}_{2}< 0,\lambda^{\prime}_{3}<0)$ in this equation 
(\ref{eqn:vf}) can be found from equations (\ref{eqn:lam}) and (\ref{eqn:ess}) 
by using the Bayes theorem again: 
\begin{eqnarray}
p(\delta | \lambda^{\prime}_{1}=\lambda_{c},\lambda^{\prime}_{2}<0,
\lambda^{\prime}_{3}<0) &=& 
\frac{p(\delta,\lambda^{\prime}_{1}=\lambda_{c},\lambda^{\prime}_{2}<0,
\lambda^{\prime}_{3}<0)}{P(\lambda^{\prime}_{1}=1,\lambda^{\prime}_{2}<0,
\lambda^{\prime}_{3}<0)}, \nonumber \\
&=& \frac{ \int^{0}_{\lambda^{\prime}_{3}}d\lambda^{\prime}_{2}
\!\int^{0}_{-\infty}d\lambda^{\prime}_{3}\,
p(\delta,\lambda^{\prime}_{1}=\lambda_{c},\lambda^{\prime}_{2},
\lambda^{\prime}_{3})} 
{ \int^{0}_{\lambda^{\prime}_{3}}d\lambda^{\prime}_{2}
\!\int^{0}_{-\infty}d\lambda^{\prime}_{3}\,
p(\lambda^{\prime}_{1}=\lambda_{c},\lambda^{\prime}_{2}.
\lambda^{\prime}_{3})} 
\end{eqnarray}

Now, the differential volume fraction $\partial F/\partial\sigma$ in equation 
(\ref{eqn:mf}) is found to be 
\begin{eqnarray}
\frac{\partial F}{\partial\sigma} &=& \frac{\partial}{\partial\sigma}
\int^{\infty}_{\delta_{c}}\!d\delta\,
p(\delta | \lambda_{c},\lambda^{\prime}_{2}<0,
\lambda^{\prime}_{3}<0),\nonumber \\
&=&-\left(\frac{\sigma}{\sigma^{2}_{\Delta}}\right)
\left[\int^{0}_{\lambda^{\prime}_{3}}d\lambda^{\prime}_{2}
\!\int^{0}_{-\infty}d\lambda^{\prime}_{3}\,
p(\lambda_{c},\lambda^{\prime}_{2},
\lambda^{\prime}_{3})\right]^{-1}\times \nonumber \\
&&\int^{0}_{\lambda^{\prime}_{3}}\! 
d\lambda^{\prime}_{2}\int^{0}_{-\infty}\! d\lambda^{\prime}_{3}\,
\left(\delta_{c}-\lambda^{\prime}_{1}-\lambda^{\prime}_{2}-
\lambda^{\prime}_{3}\right)\,
p(\delta,\lambda_{c},\lambda^{\prime}_{2},\lambda^{\prime}_{3}).
\end{eqnarray}
The logarithmic derivative of the rms density fluctuation $d\sigma/d\ln M$ 
in equation (\ref{eqn:mf}) for the case of the sharp k-space filter is also 
found to be
\begin{equation}
\label{eqn:dsi}
\frac{d\sigma}{d\ln M} = -\frac{1}{6\sigma}\Delta^{2}(\ln k_{c}),
\end{equation}
where $k_{c}$ is given in equation (\ref{eqn:den}). 

By equations (\ref{eqn:mf})-(\ref{eqn:dsi}), we evaluate the mass function 
of the low-mass halos in the mass range $10^{6}h^{-1}M_{\odot} \le M 
\le 10^{10}h^{-1}M_{\odot}$, assuming that all halos in this mass range were 
embedded in pancakes of mass $10^{11}h^{-1}M_{\odot}$ at redshift $z=2$. 
The normalization factor $A$ in equation (\ref{eqn:mf}) is determined from 
the constraint that our mass function on the mass scale 
$M=10^{10}h^{-1}M_{\odot}$ should give the same value as that of the ST 
formula which is known to agree very well with N-body simulation in the mass 
range  $M \ge 10^{10}h^{-1}M_{\odot}$.

\begin{figure}
\begin{center}
\plotone{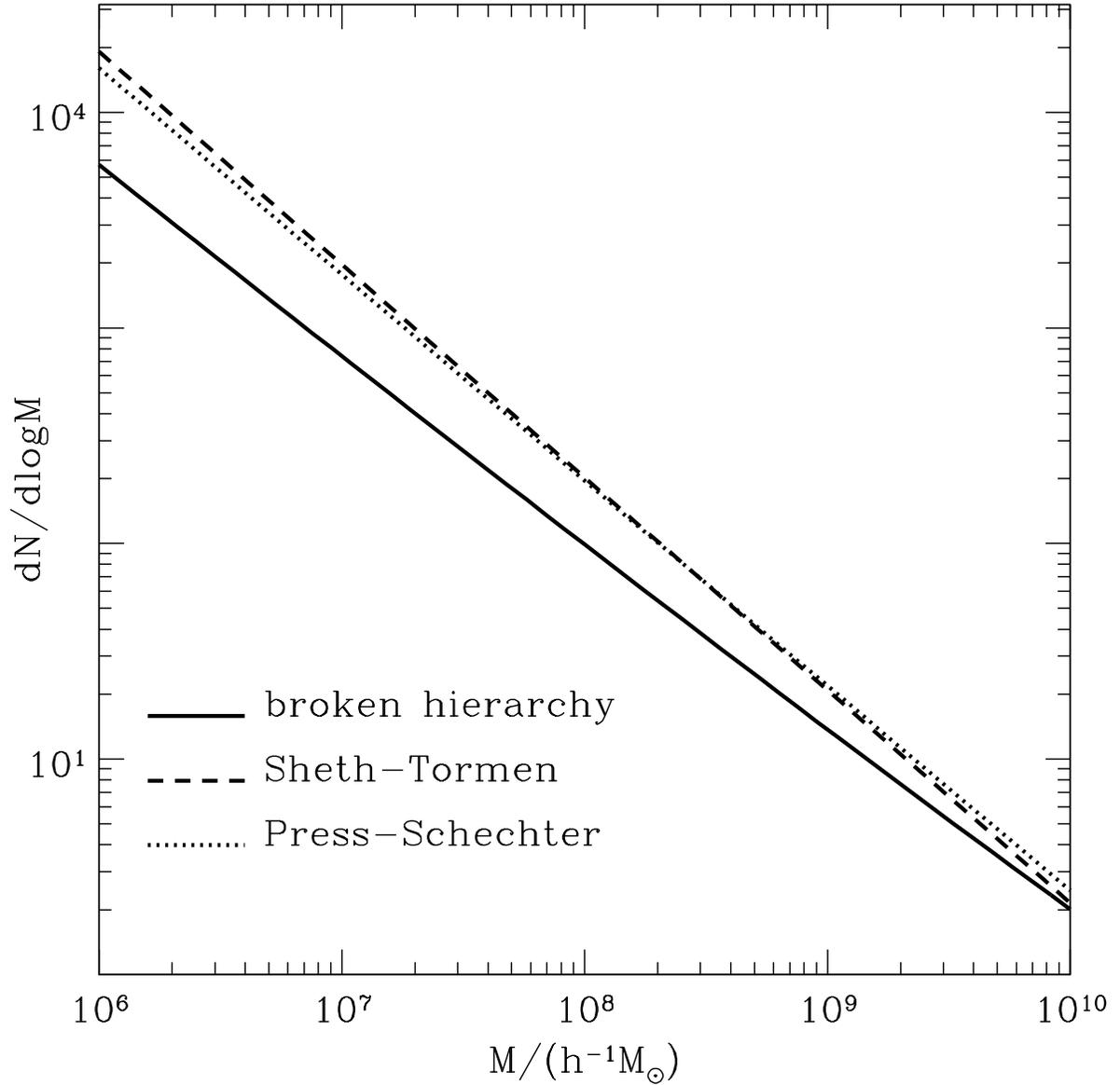}
\caption{Predictions of our (solid), the Press-Schechter (dashed) 
and the Sheth-Tormen (dotted) models for the number density of dark halos 
as a function of logarithmic mass.
\label{fig:mf}}
\end{center}
\end{figure}

Figure \ref{fig:mf} plots our result (solid), and compares it with the 
original PS (dotted) and the ST (dashed) mass functions.  
As can be seen, our model predicts less number of low-mass halos when compared 
with the PS and the ST mass functions.  That is, the formation of low-mass 
halos is suppressed by the earlier formation of massive pancakes. 
Our mass function is found to be well fitted by a power-law, 
$dN/dM \approx M^{-1.86}$, which is shallower than the PS and the ST ones, 
$dN/dM \approx M^{-2.1}$. This shallow shape of our mass function in the 
low-mass tail is consistent with the recent high-resolution simulation 
\citep{yah-etal04}. 

\section{SUMMARY AND DISCUSSION}

By taking into account the possibility that the low-mass CDM halos form 
through anti-hierarchical fragmentation of the massive pancakes, we have 
derived a new analytic mass function for the low-mass halos in the 
$\Lambda$CDM cosmology with the help of the Zel'dovich approximation and the 
Press-Schechter mass function theory. It has been shown that our mass function 
has a shallower slope in the low-mass tail and predicts maximum five times 
less abundance of dwarf galactic halos of mass $10^{6}h^{-1}M_{\odot}$ than 
the currently popular Sheth-Tormen formula. 

The concept of broken-hierarchy should modify not only the mass function but 
also the other halo statistics from the previous models that were constructed 
under the assumption that the halo formation is perfectly hierarchical. 
For instance, the two-point correlation of dwarf galactic halos would be 
different in the broken-hierarchy scenario, which in turn implies the 
mass-to-light bias on the dwarf galactic scale would be altered in accordance. 
Our future work will be in the direction of investigating the effect of 
broken-hierarchy on the halo n-point correlations and the mass-to-light 
bias as well.

Since our mass function has been derived analytically from first principles 
without introducing any fitting parameters, one may not expect it to be 
very realistic. The formation of low-mass halos should be dominated 
by complicated non-linear processes which cannot be described by using 
first principles alone. However, as it is the first attempt to model the 
broken hierarchy which can accommodate future refinements, it is concluded 
that our model will provide a useful guideline for the theoretical study 
of the effects of broken hierarchy on the structure formation.

\acknowledgments
 
This work is supported by the research grant No. R01-2005-000-10610-0 from 
the Basic Research Program of the Korea Science and Engineering Foundation.


\clearpage
\clearpage
\clearpage

\end{document}